\documentclass[twocolumn]{article}
\usepackage{imr}
\usepackage{graphicx,xcolor}
\usepackage{subcaption}
\usepackage{amsfonts,amsmath}
\usepackage{hyperref}

\begin{document}

\title{\uppercase{Parallel Metric-Based Mesh Adaptation in PETSc using ParMmg}}
\author{
    Joseph G. Wallwork$^{1,*}$
    \and Matthew G. Knepley$^2$
    \and Nicolas Barral$^{3,4}$
    \and Matthew D. Piggott$^1$
}
\date{
$^1$Department of Earth Science and Engineering, Imperial College London, London, U.K.\\
$^2$Computer Science and Engineering, State University of New York, Buffalo, NY, U.S.A.\\
$^3$University of Bordeaux, CNRS, Bordeaux INP, IMB, UMR 5251, 33400, Talence, France\\
$^4$INRIA, IMB, UMR 5251, 33400, Talence, France\\
$^*$Corresponding author. E-mail: j.wallwork16@imperial.ac.uk\\
}

\abstract{
    This research note documents the integration of the MPI-parallel metric-based mesh adaptation toolkit \emph{ParMmg} into the solver library \emph{PETSc}.
    This coupling brings robust, scalable anisotropic mesh adaptation to a wide community of PETSc users, as well as users of downstream packages.
    We demonstrate the new functionality via the solution of Poisson problems in three dimensions, with both uniform and spatially-varying right-hand sides.
}

\keywords{ mesh adaptation, anisotropy, Riemannian metric, PETSc, ParMmg, ARCHER2}

\maketitle
\thispagestyle{empty}
\pagestyle{empty}

\section{Introduction}\label{sec:intro}

Metric-based mesh adaptation is now a mature technology, which has the capacity to produce highly anisotropic, multi-scale meshes of scientifically and industrially relevant domains in two or three dimensions.
Its effectiveness as an advanced discretisation method has been demonstrated on numerous occasions, in application areas ranging from geoscience to aerospace engineering \cite{PF09,AL16}.
The core idea is to use a Riemannian metric space within the mesher, to guide the concept of mesh optimality.
Whilst an element of an adapted mesh might look distorted in Euclidean space, it is `unit' when viewed in the Riemannian space -- its edges having (close to) unit length.
A great advantage of the metric-based approach is that it allows control of element shape and orientation, as well as size.
This can be particularly beneficial for problems with strong direction-dependence and/or anisotropy.

PETSc (Portable, Extensible Toolkit for Scientific Computing) \cite{petsc-user-ref,petsc-web-page} is a widely used scientific library, written in C, which provides the linear and nonlinear solvers and associated data structures required to solve partial differential equations in parallel, among many other things.
Metric-based mesh adaptation is already available to PETSc's unstructured mesh manager, DMPlex \cite{DMPlex,LM15}, using \emph{Pragmatic}, as documented in a previous IMR research note \cite{Pragmatic}.
The software development efforts documented in this research note build upon the previous work, integrating two new toolkits -- Mmg (serial) \cite{Mmg} and ParMmg (MPI-parallel) \cite{ParMmg} -- as well as providing an interface to DMPlex, with routines for constructing, modifying and combining Riemannian metrics.
Python bindings are provided through \emph{petsc4py} \cite{petsc4py}, so that downstream packages, such as the \emph{Firedrake} finite element library \cite{Firedrake}, may take advantage of these enhancements.

The remainder of this research note is as follows.
Section \ref{sec:impl} describes the implementation of the PETSc-ParMmg coupling, including information on how it can be integrated into an existing workflow.
Section \ref{sec:demo} demonstrates its application to Poisson problems with two different choices of RHS.
Finally, an outlook on future work is made in Section \ref{sec:out}.
An appendix is provided, which includes the commands required to reproduce the experimental results.

\section{Implementation}\label{sec:impl}

\subsection{Mmg and ParMmg}\label{subsec:impl:parmmg}

\emph{Mmg} comprises a collection of tools for anisotropic metric-based adaptation of simplicial meshes \cite{Mmg}.
It includes a 2D local remesher (\emph{Mmg2d}), a 3D local remesher (\emph{Mmg3d}) and a surface remesher (\emph{Mmgs}).
The 3D local remesher has been parallelized using MPI as \emph{ParMmg} \cite{ParMmg}.
Parallel implementations do not currently exist in 2D or for surface remeshing.
Given an input mesh and a metric defined upon it, Mmg applies a sequence of operations to the mesh in order to optimize the quality of its elements, as measured by a quality functional, for which optimality means a perfectly isotropic element with edges of unit length.
In the 3D case, the operations used to optimize the mesh are ($h$-adaptive) node insertion, node deletion and face swapping, as well as ($r$-adaptive) node movement.

ParMmg performs parallel mesh adaptation iteratively, first applying Mmg3d to the mesh partition owned by each process, then repartitioning and doing the same again.
Three iterations are typically sufficient to remove dependence on the initial partition.

\subsection{PETSc DMPlex}\label{subsec:impl:plex}

In PETSc, unstructured meshes are managed by the DMPlex data structure.
The representation uses a Hasse diagram, or directed acyclic graph \cite{DMPlex}, whereby all entities are treated equally as vertices of the graph.
This enables algorithms to be written independently of mesh properties such as dimension and cell type~\cite{KL15}.
DMPlex supports all of the operations required to handle meshing, such as generation, partitioning and distribution, creating of missing edges and faces, regular refinement, extrusion, traversal, selection, manipulation and I/O~\cite{HK20}.
This work adds Riemannian metric utilities, to support metric-based adaptation.

\subsection{Coupling}\label{subsec:impl:coupling}

\paragraph{Overview}

Given a linear simplicial mesh described using a DMPlex, a Riemannian metric field may be defined in the finite element context as a piece-wise linear and continuous ($\mathbb P1$) field, with its degrees of freedom (DoFs) at mesh vertices.
The values at these DoFs can be encapsulated in a PETSc Vec.
The new functionality introduced by this paper provides routines for the construction (e.g.~\href{https://petsc.org/main/docs/manualpages/DMPLEX/DMPlexMetricCreate.html}{\texttt{DMPlexMetricCreate}}) and modification (e.g.~\href{https://petsc.org/main/docs/manualpages/DMPLEX/DMPlexMetricEnforceSPD.html}{\texttt{DMPlexMetricEnforceSPD}}) of such objects, as well as for combining different instances (e.g.~\href{https://petsc.org/main/docs/manualpages/DMPLEX/DMPlexMetricIntersection.html}{\texttt{DMPlexMetricIntersection}}).
We make use of the $L^p$ normalization techniques introduced by \cite{LA11b}, which both ensure that resulting adapted meshes have the desired complexity and enable them to be appropriately multi-scale (\href{https://petsc.org/main/docs/manualpages/DMPLEX/DMPlexMetricNormalize.html}{\texttt{DMPlexMetricNormalize}}).

Given a metric constructed using these routines, \href{https://petsc.org/main/docs/manualpages/DM/DMAdaptMetric.html}{\texttt{DMAdaptMetric}} provides the interface so that ParMmg can be used to drive the mesh adaptation process.
The Pragmatic interface described in \cite{Pragmatic} is also maintained.
When invoked, the Vec containing the metric data is converted to an array of \href{https://petsc.org/main/docs/manualpages/Sys/PetscScalar.html}{\texttt{PetscScalar}}s (i.e.~\texttt{double}s in real mode) and passed to the appropriate remesher.
In addition to the metric operations above, Mmg and ParMmg have in-built metric gradation functionality, which ensures that two adjacent edges in the mesh do not differ by too large a ratio.

\paragraph{Hessian Metric}

An appropriately scaled Hessian is a common basis for the construction of an error metric, since it incorporates anisotropic information related to the curvature of the solution and is related to interpolation error \cite{LA11b}.
Its symmetry implies an orthogonal eigendecomposition at each DoF, which can be computed straightforwardly.
A metric can then be constructed by simply taking the eigenvalues in modulus.
This is justified because error magnitudes are typically more important than their signs.

\paragraph{Integration into an Existing Workflow}

PETSc's nonlinear solver module, SNES, has inbuilt capacity for mesh adaptation.
It allows the user to take a fixed point iteration approach, whereby the PDE is solved, a Hessian is recovered from its solution and the mesh is adapted w.r.t.~the Hessian at each iteration.
For a sufficiently well-behaved steady-state PDE and appropriate metric parameters, the process typically converges in a small number of iterations.
The timestepping object, TS, has a similar facility.

The Hessian recovery step is implemented using two applications of the Cl\'ement interpolant \cite{Cle75} -- one to recover the gradient and another for the Hessian.
The parallel computation of Cl\'ement interpolation requires that the \emph{star} of every vertex is available, i.e.~the set of all cells containing the vertex.
DMPlex is able to generate an arbitrary mesh overlap in parallel using \href{https://petsc.org/main/docs/manualpages/DMPLEX/DMPlexDistributeOverlap.html}{\texttt{DMPlexDistributeOverlap}}.
Then we project the average gradient over the star into $\mathbb P1$.
A metric is constructed from the Hessian using $L^p$-normalization, with the default $p=1$ allowing multi-scale meshing.

If the executable for a particular run involving nonlinear solves is called \texttt{ex} then the application of metric-based mesh adaptation with \texttt{n} fixed point iterations can be as simple as using the command line options
\begin{verbatim}
./ex -dm_adaptor mmg -snes_adapt_sequence <n>.
\end{verbatim}
To run in parallel with \texttt{np} MPI processes, simply call
\begin{verbatim}
mpiexec -np <np> ./ex -dm_adaptor parmmg \
                      -snes_adapt_sequence <n>.
\end{verbatim}

\section{Demonstration}\label{sec:demo}

To demonstrate the coupling between PETSc and ParMmg, we solve the Poisson equation in parallel on adaptive meshes of the unit cube, $\Omega=[0,1]^d$, $d=3$:
\begin{equation}\label{eq:demo:poisson}
    \Delta u=f\quad\text{in}\:\Omega,\qquad
    u=g\quad\text{on}\:\partial\Omega,
\end{equation}
with solution $u\in H^2(\Omega)$, forcing $f\in L^2(\Omega)$ and Dirichlet condition defined by $g\in L^2(\partial\Omega)$.
By choosing different $f$ and $g$, we may construct manufactured solutions with different features.
We discretize by seeking weak solutions in order $k\in\mathbb N$ Lagrange space, $\mathbb Pk\subset H^1(\Omega)$.
The resulting linear systems are solved using conjugate gradients, preconditioned with SOR.

\subsection{Hessian of a Quadratic Function}\label{subsec:demo:uniform}

Suppose we seek the quadratic manufactured solution,
\begin{equation}\label{eq:demo:uniform:sol}
    u_1(\mathbf x):=\tfrac23\mathbf x\cdot\mathbf x,\quad\mathbf x:=(x,y,z).
\end{equation}
This can be achieved with $f\equiv4$ and appropriate Dirichlet conditions.
Three fixed point iterations are performed, to reduce dependence on the initial mesh.

\begin{figure}[h]
    \centering
    \begin{subfigure}{0.24\textwidth}
        \centering
        \includegraphics[width=\textwidth]{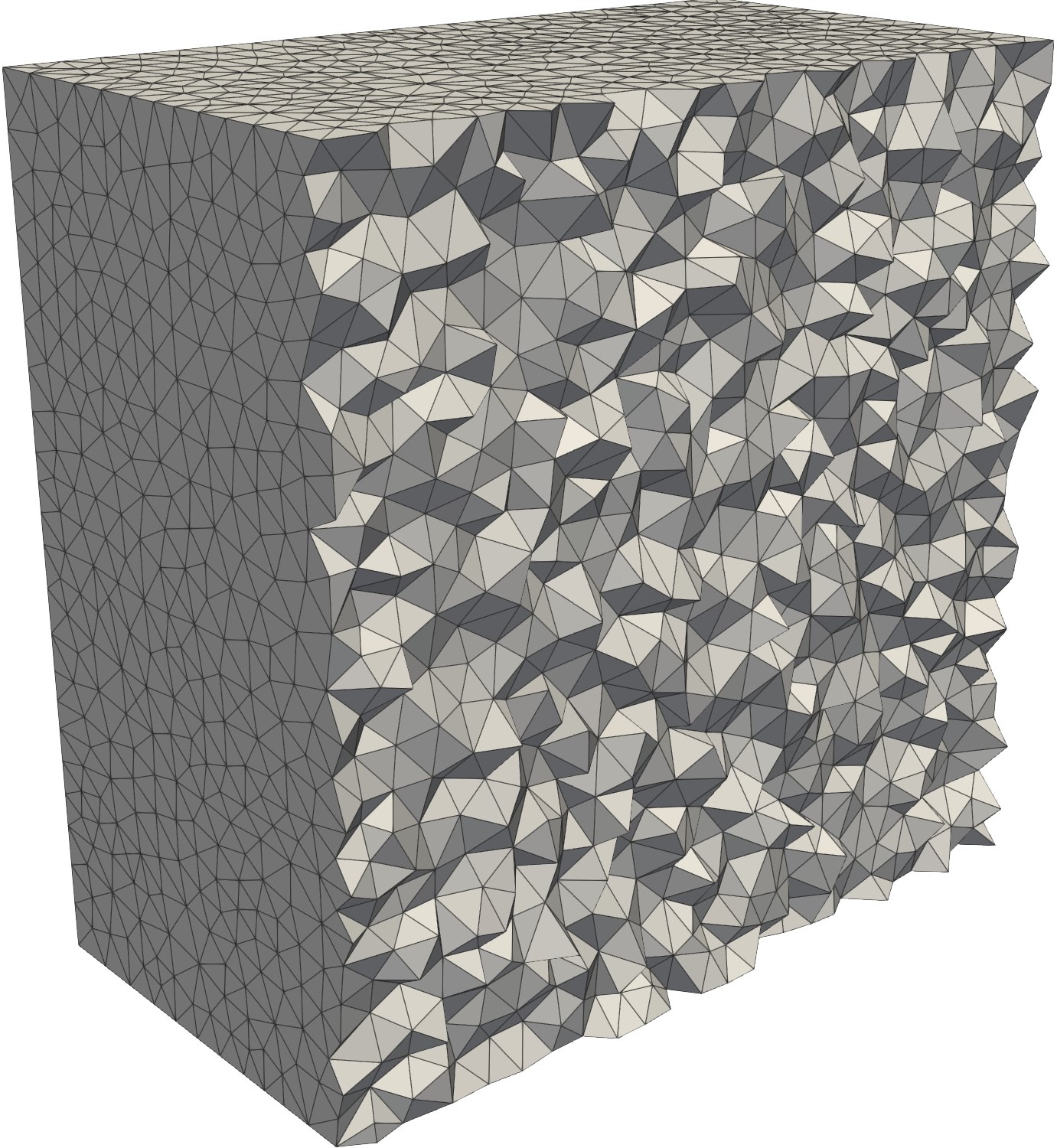}
    \end{subfigure}
    \begin{subfigure}{0.2\textwidth}
        \centering
        \includegraphics[width=\textwidth, trim={20 20 20 20}, clip]{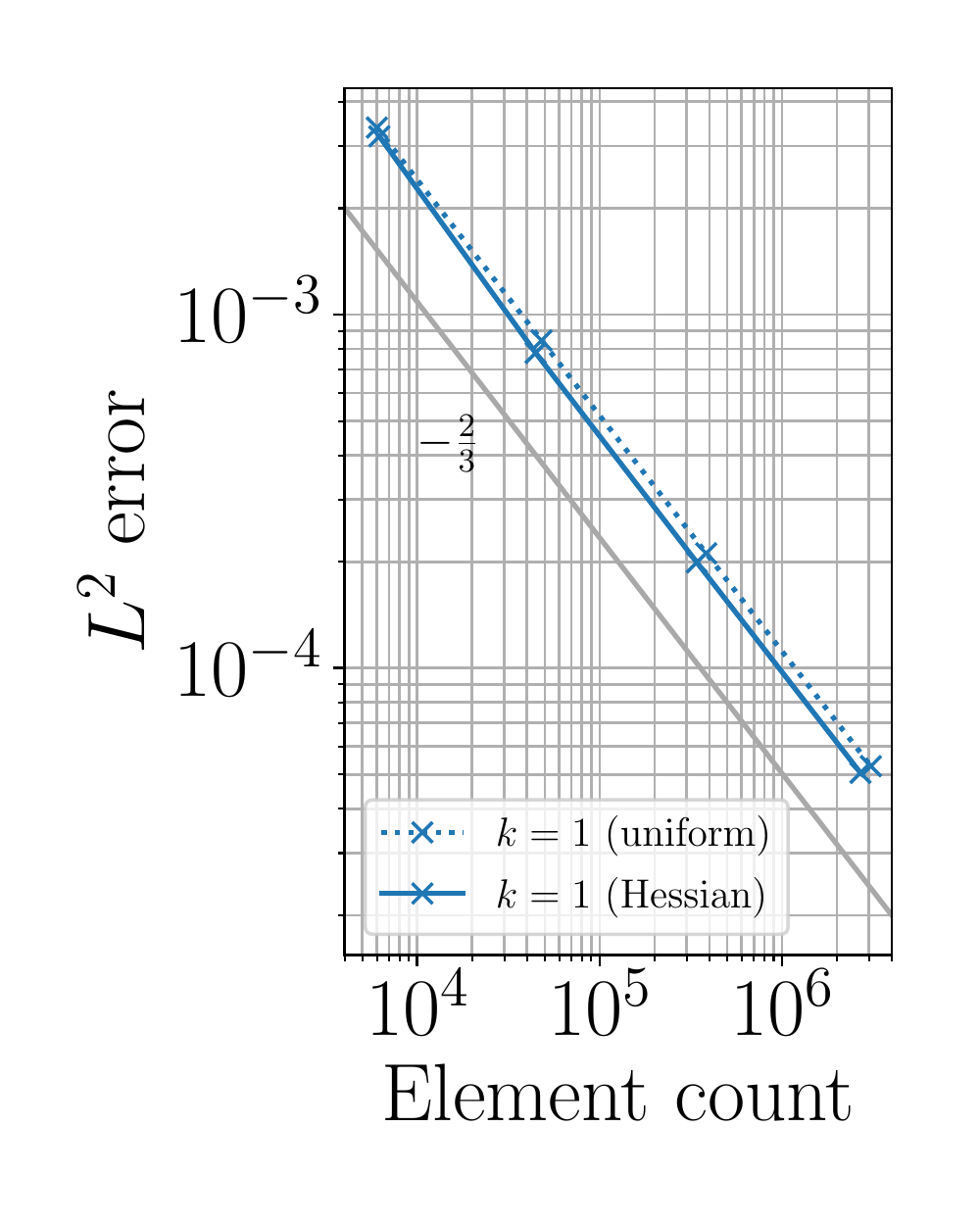}
    \end{subfigure}
    \caption{
        Left: mesh generated by adapting w.r.t.~the Hessian of the weak solution of (\ref{eq:demo:poisson}) with RHS $f\equiv4$.
        Right: $L^2$ error convergence analysis.
        Mesh statistics: 88,362 elements, 18,534 vertices, max. aspect ratio 5.2981, mean aspect ratio 1.4051, std. dev. 0.1898.
    }
    \label{fig:demo:uniform}
\end{figure}

The mesh plot in Figure \ref{fig:demo:uniform} verifies that the uniform Hessian can be recovered and that the resulting adapted mesh is indeed isotropic.
Some elements are moderately anisotropic (with aspect ratios greater than two), although this is only true for 0.6\% of the elements.
The presence of some slightly anisotropic elements is unavoidable in tetrahedral mesh adaptation since unit tetrahedra cannot tile space in general.

By increasing target metric complexity, we may produce increasingly refined meshes and thereby perform convergence analysis.
The right-hand plot in Figure \ref{fig:demo:uniform} demonstrates the $(k+1)/d$ convergence rate in the $L^2$ norm for degree $k=1$.
This corresponds to the expected $\mathcal O(h^{k+1})$ convergence on isotropic meshes, which is matched by standard uniform refinement.

\subsection{Interface Capturing}\label{subsec:demo:interface}

Next, we demonstrate the capability to capture interfaces.
We seek the manufactured solution,
\begin{equation}\label{eq:demo:interface:sol}
    u_2(\mathbf x):=\tanh(\alpha(r^2-(\mathbf x-\mathbf x_0)\cdot(\mathbf x-\mathbf x_0))),
\end{equation}
which is a smooth approximation of an indicator function for a sphere of radius $r>0$, centered on $\mathbf x_0:=(0.5,0.5,0.5)$.
The value of $\alpha>0$ determines the sharpness of the interface is between the exterior and the interior of the sphere.
Take $\alpha=500$ and $r=0.15$.
Differentiation implies the forcing
\begin{equation}\label{eq:demo:interface:forcing}
    f(\mathbf x):=2\alpha(u_2^2(\mathbf x)-1)(4\alpha(\mathbf x-\mathbf x_0)\cdot(\mathbf x-\mathbf x_0)\,u_2(\mathbf x)+d).
\end{equation}

\begin{figure}[t]
    \centering
    \begin{subfigure}{0.24\textwidth}
        \centering
        \includegraphics[width=\textwidth]{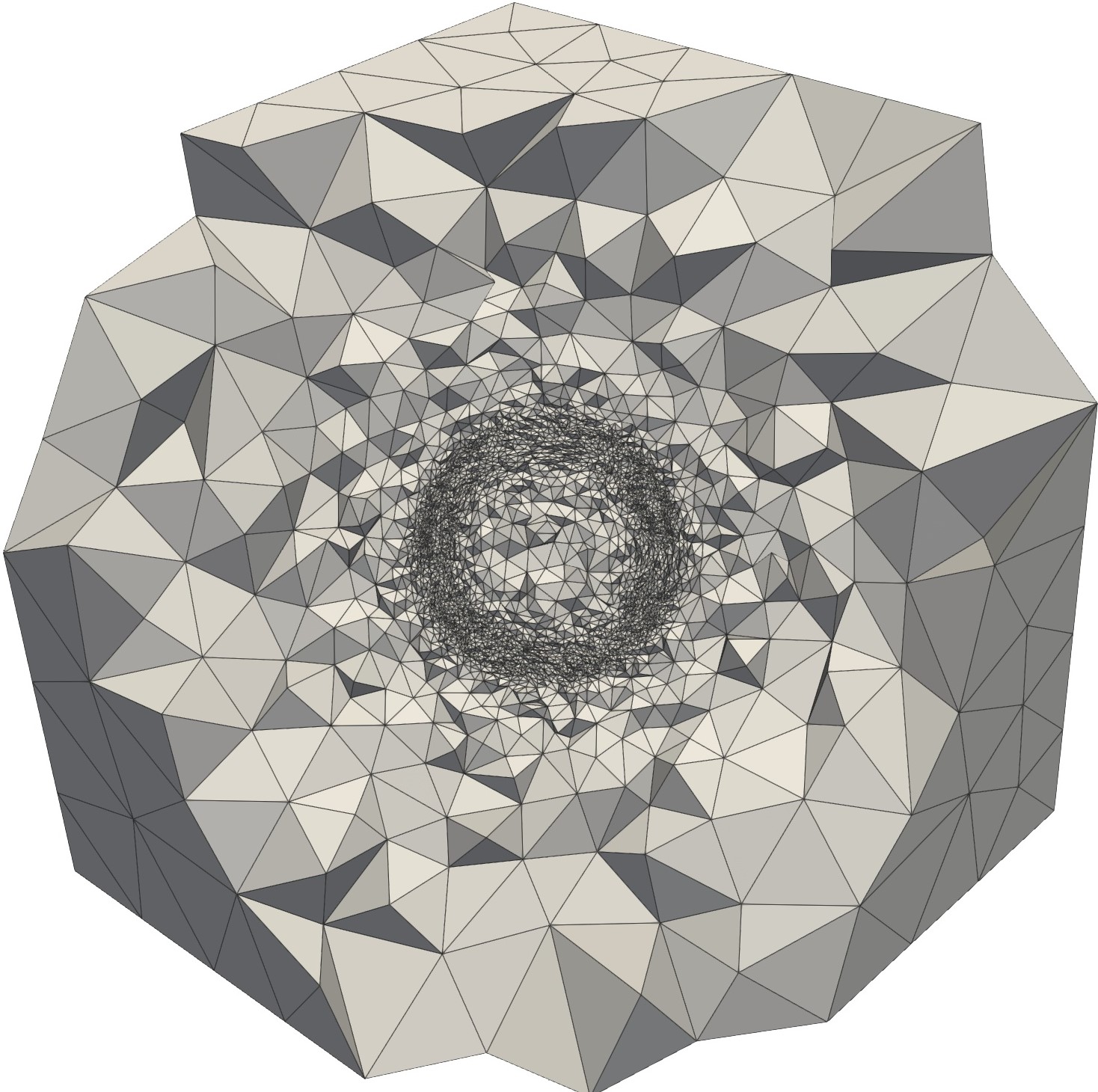}
    \end{subfigure}
    \begin{subfigure}{0.2\textwidth}
        \centering
        \includegraphics[width=\textwidth, trim={20 20 20 20}, clip]{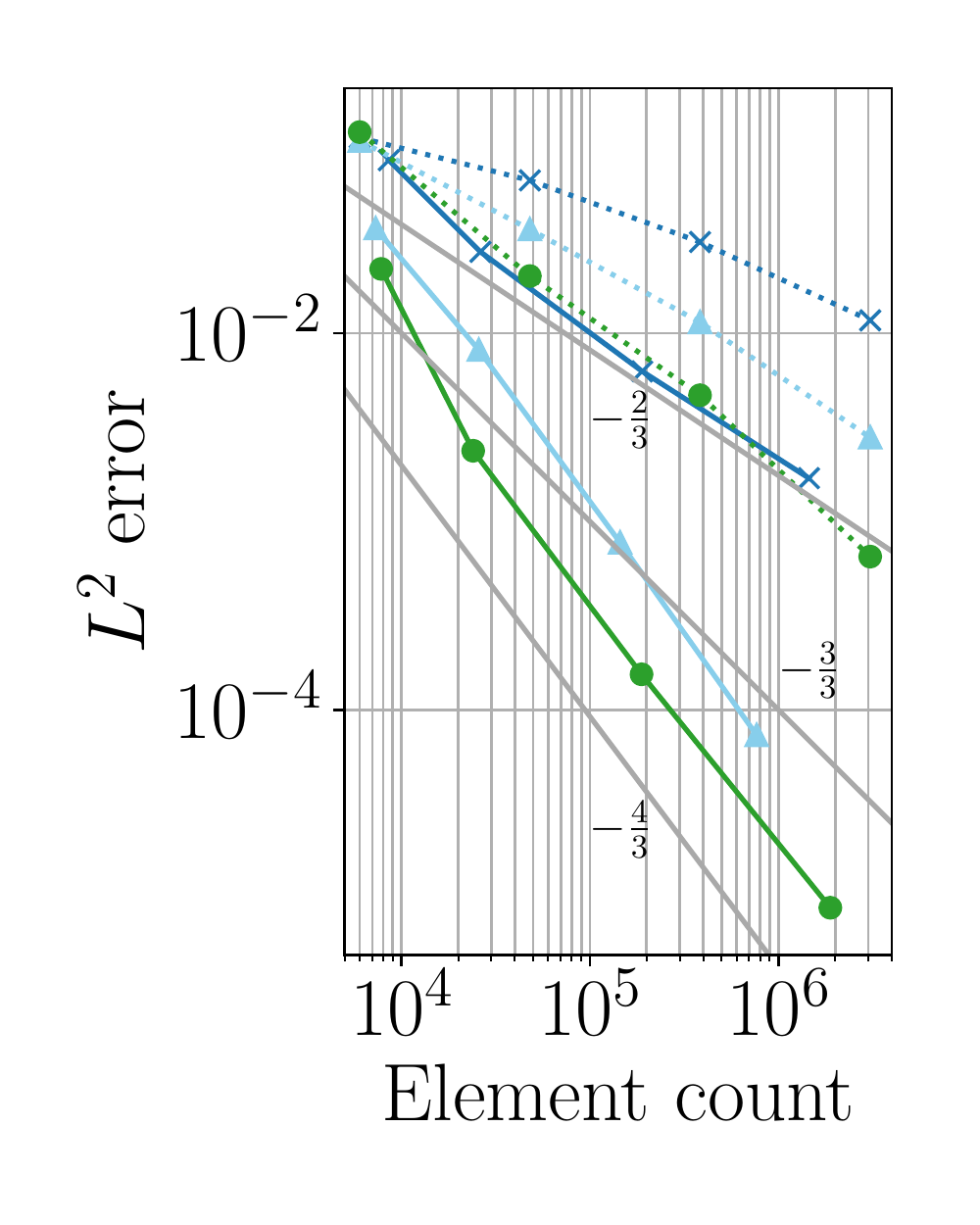}
        \includegraphics[width=\textwidth]{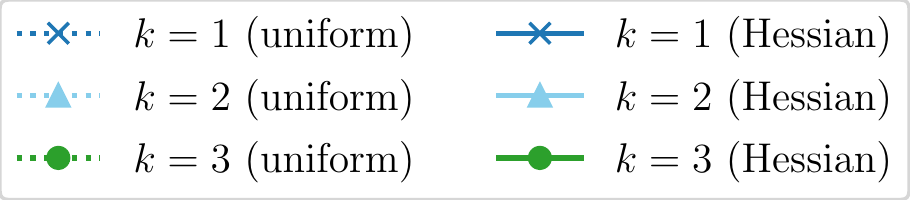}
    \end{subfigure}
    \caption{
        Left: mesh generated by adapting w.r.t.~the Hessian of the weak solution of (\ref{eq:demo:poisson}) with RHS (\ref{eq:demo:interface:forcing}).
        Right: $L^2$ error convergence analysis.
        Mesh statistics: 595,594 elements, 105,960 vertices, max. aspect ratio 21.0797, mean aspect ratio 1.6564, std. dev. 0.3907, min./max. volumes: $3.7504\times10^{-10}$/$1.3294\times10^{-3}$.
    }
    \label{fig:demo:interface}
\end{figure}

The left-hand plot in Figure \ref{fig:demo:interface} shows the adapted mesh for target complexity 32,000.
High resolution is used to capture the interface, with some fairly anisotropic elements.
Notably coarse resolution is used near the domain boundaries.
The caption indicates that we have a multi-scale mesh, whose element volumes span eight orders of magnitude.
Elements are generally more anisotropic for this test case, with 14.8\% having aspect ratio greater than two.

The Hessian-based curves in Figure \ref{fig:demo:interface} exhibit the same $(k+1)/d$ convergence rate as in Subsection \ref{subsec:demo:uniform}.
However, uniform refinement does not attain the optimal rate for this particular problem.
As such, we demonstrate that, together, PETSc and ParMmg are able to achieve optimal $L^2$ convergence for a Poisson problem with a spatially varying RHS, for which standard uniform refinement is unable to do this.

\section{Outlook}\label{sec:out}

The developments described herein comprise the first phase of a project that aims to bring metric-based mesh adaptation to the user communities of both PETSc and the downstream Firedrake finite element library \cite{Firedrake}.
The next phase is to facilitate mesh adaptation in Firedrake, by implementing consistent mesh-to-mesh interpolation schemes in parallel, perhaps based on those currently in PETSc.
The project focuses on \emph{goal-oriented} adaptation, which enables the accurate approximation of diagnostic quantities using relatively few DoFs overall.
Such technologies are currently being implemented as part of this project, based on previous proof-of-concept work including \cite{imr_paper}.
The final phase will include performing benchmarking experiments and writing extensive documentation.

\section*{Appendix}

The numerical experiments conducted in this paper use one of PETSc's standard test cases, which can be found at \texttt{\$PETSC\_DIR/src/snes/tutorials/ex12.c} in the source code \cite{petsc-web-page}.
Having compiled this example, adaptation to a uniform metric (as in Subsection \ref{subsec:demo:uniform}) can be done by running
\begin{verbatim}
mpiexec -np <np> ./ex12 -run_type full
    -dm_plex_dim 3 -dm_distribute
    -dm_plex_box_faces 10,10,10
    -bc_type dirichlet -petscspace_degree <k>
    -variable_coefficient none
    -ksp_type cg -pc_type sor 
    -snes_adapt_sequence <n>
    -adaptor_target_num <target>
    -dm_plex_metric_h_max 0.5
    -dm_adaptor parmmg
\end{verbatim}
Here \texttt{np} and \texttt{n} again encode the number of processors and the number of fixed point iterations in the adaptation loop, whilst \texttt{k} and \texttt{target} encode the polynomial degree for the finite element space and the target metric complexity.
The convergence analysis experiments may be performed by varying these values.
For the interface capturing example in Subsection \ref{subsec:demo:interface}, simply replace with
\begin{verbatim}
    -variable_coefficient ball
\end{verbatim}
in the appropriate line above.

\begin{center}
    \subsectionfont Acknowledgments
\end{center}

This work was funded under the embedded CSE program of the ARCHER2 UK National Supercomputing Service (\url{http://www.archer2.ac.uk}).
Support is also acknowledged from EPSRC under grant EP/R029423/1.

\bibliography{ref}

\end{document}